\documentclass[manuscript,screen,nonacm]{acmart}
\AtBeginDocument{%
  }

\setcopyright{acmlicensed} 
\copyrightyear{2027} 
\acmYear{2027} 
\acmDOI{XXXXXXX.XXXXXXX} 
\acmConference[FAccT '27]{ACM Conference on Fairness, Accountability, and Transparency}{June 21--24, 2027}{Porto, Portugal} 
\acmISBN{978-1-4503-XXXX-X/2027/06} 

\usepackage[capitalize,noabbrev]{cleveref}

\usepackage{xspace}

\newcommand{\Trak}{\textsc{Trak}\xspace}
\newcommand{\FactGraSS}{\textsc{FactGraSS}\xspace}
\newcommand{\LoGra}{\textsc{LoGra}\xspace}
\newcommand{\GradCos}{\textsc{Grad-Cos}\xspace}

\Crefname{equation}{Eq.\!}{Eqs.\!}
\Crefname{table}{Table}{Tables}
\Crefname{figure}{Figure}{Figures}

\newcommand{\CLAP}{CLAP\xspace}

\newcommand{\MAEST}{MAEST\xspace}

\newcommand{\Global}{\textit{global}\xspace}
\newcommand{\Group}{\textit{group}\xspace}
\newcommand{\Specific}{\textit{specific}\xspace}

\usepackage{url}            
\usepackage{multirow}
\usepackage{tabularx}
\usepackage{nicefrac}       
\usepackage{microtype}      
\usepackage{enumitem}       
\usepackage{float}          
\usepackage{wrapfig}        
\usepackage{placeins}       
\usepackage{pgfplots}
\pgfplotsset{compat=1.18}
\usepackage{subcaption}
\usepackage{tikz}
\usepgfplotslibrary{groupplots}

\title{Functional Dependency beyond Similarity: Hierarchical Decomposition of Training Data Attribution in Music Generative Models}

\author{Changheon Han}
\affiliation{%
  \institution{Chalmers University of Technology and University of Gothenburg}
  \department{Department of Computer Science and Engineering}
  \city{Gothenburg}
  \country{Sweden}}
\email{changheon.han@chalmers.se}

\author{Hugh Xuechen Liu}
\affiliation{%
  \institution{Chalmers University of Technology and University of Gothenburg}
  \department{Department of Computer Science and Engineering}
  \city{Gothenburg}
  \country{Sweden}}
\email{xuechen@chalmers.se}

\author{Ashkan Panahi}
\affiliation{%
  \institution{Chalmers University of Technology and University of Gothenburg}
  \department{Department of Computer Science and Engineering}
  \city{Gothenburg}
  \country{Sweden}}
\email{ashkan.panahi@chalmers.se}

\author{K{\i}van\c{c} Tatar}
\correspondingauthor
\affiliation{%
  \institution{Chalmers University of Technology and University of Gothenburg}
  \department{Department of Computer Science and Engineering}
  \city{Gothenburg}
  \country{Sweden}}
\email{tatar@chalmers.se}

\renewcommand{\shortauthors}{Han et al.}

\begin{document}

\begin{abstract}
Copyright analysis focuses on the similarity between two works. However, generative AI models learn from training data through functional relations beyond similarity.
Training Data Attribution (TDA) estimates the influence of each training instance on a generated output and may serve as a basis for remuneration.
However, prior studies find global influence, where a few training instances rank as the most influential for almost every generation.
Used directly, TDA scores would distribute part of every remuneration to these instances regardless of what was generated.
To address this, we propose a new approach in which TDA scores are decomposed into three levels of influence, namely \textit{global} indicating the influence of a training data to a trained model, \textit{group} for its influence on a group of generated outputs (that could be clustered together), and \textit{specific} for its influence on a single generated output. Our study finds that, across several music generative models and TDA methods, \textit{global}- and \textit{group}-level influence is non-negligible in every case.
Based on these observations, we propose the term \textbf{functional dependency} for the relation of training data to a trained model or to its outputs.
This notion both encompasses and distinguishes dependencies of different scopes, allowing TDA scores to be interpreted as distinct forms of contribution in legal, economic and ethical discussions on remuneration.\footnote{Code: https://github.com/chh-han/Functional-Dependency-beyond-Similarity}
\end{abstract}


\keywords{Generative AI, Copyright, Training Data Attribution, Functional Dependency, Remuneration}

\maketitle

\section{Introduction}

While copyright investigates similarity between creative works, generative AI models~\cite{agostinelli2023musiclm,copet2023musicgen,zhang2025inspiremusic,garcia2023vampnet,chen2024musicldm,melechovsky2024mustango} bring new challenges to conventional copyright~\cite{umg2024suno,umg2024udio,usco2025part3,dornis2025copyright,kretschmer2025uk,geiger2024elaborating} and royalty allocations~\cite{senftleben2023generative,geiger2024forgotten,samuelson2026feasibility,axhamn2025ecl,ep2026copyright}.
With generative AI models, a creative work can be utilized beyond its similarity, through functional relations to other creative works from the process of training a model to generating outputs.
Therefore, a straightforward remuneration based on the similarity between data and a generated output may not fully reflect how the data actually contributes~\cite{wuhrl2026attribution,wei2025interrogating,deng2023computational}.

One technical approach to analyze these functional relations between training data, a model, and generated output is Training Data Attribution (TDA), which estimates the influence of each training instance\footnote{Following the TDA studies~\cite{hanawa2021evaluation,hammoudeh2022identifying,hammoudeh2024survey}, we refer to each unit of the training data that receives a score as a training instance.} on a model's generated output~\cite{koh2017understanding,pruthi2020estimating,park2023trak,hammoudeh2024survey}.
It is considered one possible basis for such a remuneration mechanism~\cite{deng2023computational}.
However, linking these estimates to transparent remuneration is not straightforward~\cite{wuhrl2026attribution}.
The interpretation of TDA scores as contributions for remuneration requires interdisciplinary discussion, involving legal, economic, and ethical considerations.
Current TDA methods typically provide a single score for each training instance and each generated output, not providing enough information for the interpretation of the influence of training instances for different discussions of remuneration~\cite{han2026aria}.
Our work aims to facilitate this discussion while quantifying how influence is distributed across different levels of dependency.

Previous studies find global influence, where a few training instances keep appearing as the most influential training instances for almost any generation~\cite{barshan2020relatif,hammoudeh2022identifying,hammoudeh2024survey}.
It has been attributed to training instances with a large gradient norm, such as outliers and training instances that the model predicts poorly during the training process~\cite{hanawa2021evaluation,hammoudeh2022identifying,hammoudeh2024survey}.
This global influence relates to the model's global behavior rather than to any one generated output~\cite{barshan2020relatif,hammoudeh2022identifying}.
The score of a training instance from TDA for one generated output may then mix this global influence with influence particular to that generated output.
If such scores were used for remuneration, part of the remuneration would keep going to the same training instances for almost any generation.
Even without this global influence, the residual of a score may still contain a part shared by a group of generated outputs.
For the remuneration to be transparent about which influence it pays for, TDA scores need to be examined at a resolution that distinguishes influence particular to one generated output from influence shared by a group of generated outputs or global influence.

In this work, we decompose TDA scores acquired from a model into three levels of dependency, which are \Global, \Group and \Specific, by subtracting the averaged influence at each level over its corresponding scope.
The \Global level is the average of a training instance's scores over all generated outputs.
It represents the dependency of the model itself on the training instance, rather than that of any specific generated output.
The \Group level is the average over each group of generated outputs, where generated outputs are grouped by a criterion such as $k$-means clustering with each column of TDA scores as a feature vector after the \Global level is subtracted.
This level represents a style as the model itself defines it, shared among generated outputs whose TDA scores follow similar patterns across training instances, rather than a genre or style defined from the data.
Lastly, the \Specific level is what remains after the \Global and \Group levels are removed.
It represents the dependency of one specific generated output on the training instance.
The decomposition thus splits each score into three levels that differ in scope, namely the model's global behavior, a style within the model and a specific generated output.

We apply the decomposition to two pretrained autoregressive music generative models in two sizes each, MusicGen-S and MusicGen-L~\cite{copet2023musicgen} and InspireMusic-Base and InspireMusic-1.5B~\cite{zhang2025inspiremusic}. Each model is fine-tuned on the MTG-Jamendo dataset~\cite{bogdanov2019mtgjamendo} and tested with six TDA methods and three variants~\cite{charpiat2019input,hanawa2021evaluation,pruthi2020estimating,park2023trak,choe2024logra,hu2026grass}.
We evaluate whether the different levels of influence in the decomposition have a counterfactual effect on the model within their scopes~\cite{park2023trak,bae2024training,chen2026mechanistic}. For each level, we remove the training instances that score highest in that level and fine-tune the pretrained model again to examine whether this affects the model’s behavior in the corresponding scope more than removing a random set of the same size.
We observe in every experimental case that the TDA scores without decomposition contain a part common to all generated outputs and a part shared by a group of generated outputs.
This means that \Specific-level influence is layered beneath influence shared at the \Group and \Global levels.
This observation suggests that any attempt to use TDA scores transparently as a basis for remuneration needs to account for the hierarchical nature of influence. Based on these observations, we define \textbf{functional dependency} as the relations through which a trained model, or a generated output from a trained model, depends on training data.
This notion both encompasses and distinguishes dependencies of different scopes, from the model itself to a single generated output.
Inspired by the \textit{structural} perspective in the discourse around LLMs and copyright~\cite{wei2025interrogating}, we use it to analyze the influence of each training instance and to support a remuneration mechanism. Our contributions are as follows.

\begin{itemize}
  \item We define functional dependency as the dependency of a trained model, or a generated output from a trained model, on training data.
  \item We propose a decomposition of TDA scores into the \Global, \Group and \Specific levels, which splits each standardized score exactly into parts of different scopes representing different dependencies.
  \item Our decomposition can be applied to any TDA method, since it leverages only the TDA scores obtained as a result.
  \item We show across four music generative models and six TDA methods and three variants that TDA scores contain \Global- and \Group-level influence in every case.
  \item Our evaluation indicates that each level has an effect on the model's behavior within its scope.
\end{itemize}

\section{Background}
\label{sec:background}

\subsection{Remuneration for Generative AI and Functional Dependency as Structural Perspective}
\label{sec:royalty-allocation}

\subsubsection{Similarity and Functional Dependency}
While copyright infringement focuses on substantial similarity~\cite{lee2023talkin}, generative AI brings new challenges for remuneration as it generates outputs influenced by many training instances rather than direct copying.
Furthermore, substantial similarity is challenging to specify algorithmically~\cite{wei2025interrogating}, and it is insufficient for capturing the functional dependencies between training data and generated outputs through generative AI~\cite{brando2025technological}.
Prior work on copyright and LLMs distinguishes a \textit{behavioral} perspective, which evaluates whether generated outputs are substantially similar to training data, from a \textit{structural} perspective, which focuses on how a model is trained~\cite{wei2025interrogating}.
Our analysis of functional dependencies is inspired by the structural perspective, since it examines what a trained model depends on rather than the similarity of its outputs.

Legal scholars have proposed that providers of generative AI systems may owe a general payment for training on copyrighted works~\cite{geiger2024forgotten,geiger2024elaborating} or a lump sum based on their revenue~\cite{senftleben2023generative}.
In recent EU policy discussions, author societies have proposed remuneration for authors when their works have been used and have influenced a generated output~\cite{eprs2026outputs}. These discussions also note that the factors establishing such influence of the original works on the output remain to be specified~\cite{eprs2026outputs}.
An attribution method that measures the influence of each training instance on a generated output is one way to establish such influence.
Remuneration based on such influence thus requires understanding the functional dependency between training data and generated outputs, which persists in a model trained on the data~\cite{brando2025technological}.

Prior work on TDA also supports the view that the influence of training instances is not aligned with their similarity to the generated output, and that some training instances that are similar to the generated output even have a negative impact~\cite{koh2017understanding,deng2023computational}.
In image classification, the most influential training instances for a test image are not necessarily of the same class as the test image~\cite{koh2017understanding}.
In music generation, the similarity of training instances to a generated output aligns poorly with their influence on average, although the two agree on the most influential training instances, and some training instances that are similar to the generated output have a negative impact~\cite{deng2023computational}. These findings suggest that training instances relate to generated outputs through functional dependencies beyond similarity.

\subsubsection{Attribution in Music Generation}
Recent work on attribution in music generation has begun to relate a generated output to the musical works behind it.
Generated outputs have been compared with training instances through the similarity of embeddings from a separate audio encoder, with listening tests as a check~\cite{barnett2024exploring}.
This approach captures the aspects of musical similarity that the encoder is sensitive to, rather than the causal influence of training instances on the generated output~\cite{han2026aria}.
Another approach proposes attribution-based music generative systems, where users select reference songs at generation time and the system attributes the generated output to these selected songs~\cite{morreale2025attribution}.
A royalty framework proposes to apply TDA to a music generative model and to distribute royalties according to the resulting scores~\cite{deng2023computational}.

\subsection{Training Data Attribution}
\label{sec:background-tda}

\subsubsection{Influence Estimation and Evaluation}
TDA is one possible approach to estimating the functional dependency between training instances and generated outputs, estimating the causal relation between them.
It estimates this dependency as the influence of each training instance on a generated output, through the counterfactual effects that quantify how a generated output would change if a training instance had not been in the training data~\cite{koh2017understanding,hammoudeh2024survey}.
Estimating these effects directly requires retraining the model once for each training instance, which is called leave-one-out retraining.
This is infeasible for large training datasets or models due to the computational cost~\cite{koh2017understanding,choi2025large}.
Gradient-based TDA methods estimate the influence of training instances without retraining, by comparing the gradient of the loss on a training instance with the gradient of a generated output~\cite{charpiat2019input,hanawa2021evaluation,pruthi2020estimating,park2023trak,choe2024logra,hu2026grass}. As a result of TDA, we obtain one score for every training instance with respect to a specific generated output. The TDA scores are usually verified by comparing them with the causal effects measured by retraining the model after removing the training instances~\cite{park2023trak,deng2024dattri}.
The causal effects measured by removing a single training instance can be noisy and vary across retraining runs~\cite{hammoudeh2024survey,basu2021influence,bae2024training}. Evaluations therefore also remove a set of high-scoring training instances at once and retrain~\cite{park2023trak,basu2021influence,bae2024training}.
The retrained model is then compared with the model retrained after randomly removing the same number of training instances~\cite{chen2026mechanistic,bae2024training}.
Through this comparison, it can be observed whether the high-scoring training instances identified by the TDA method have a larger effect on the model's behavior than a random set.

\begin{figure}[t]
  \centering
  \includegraphics[width=0.8\linewidth]{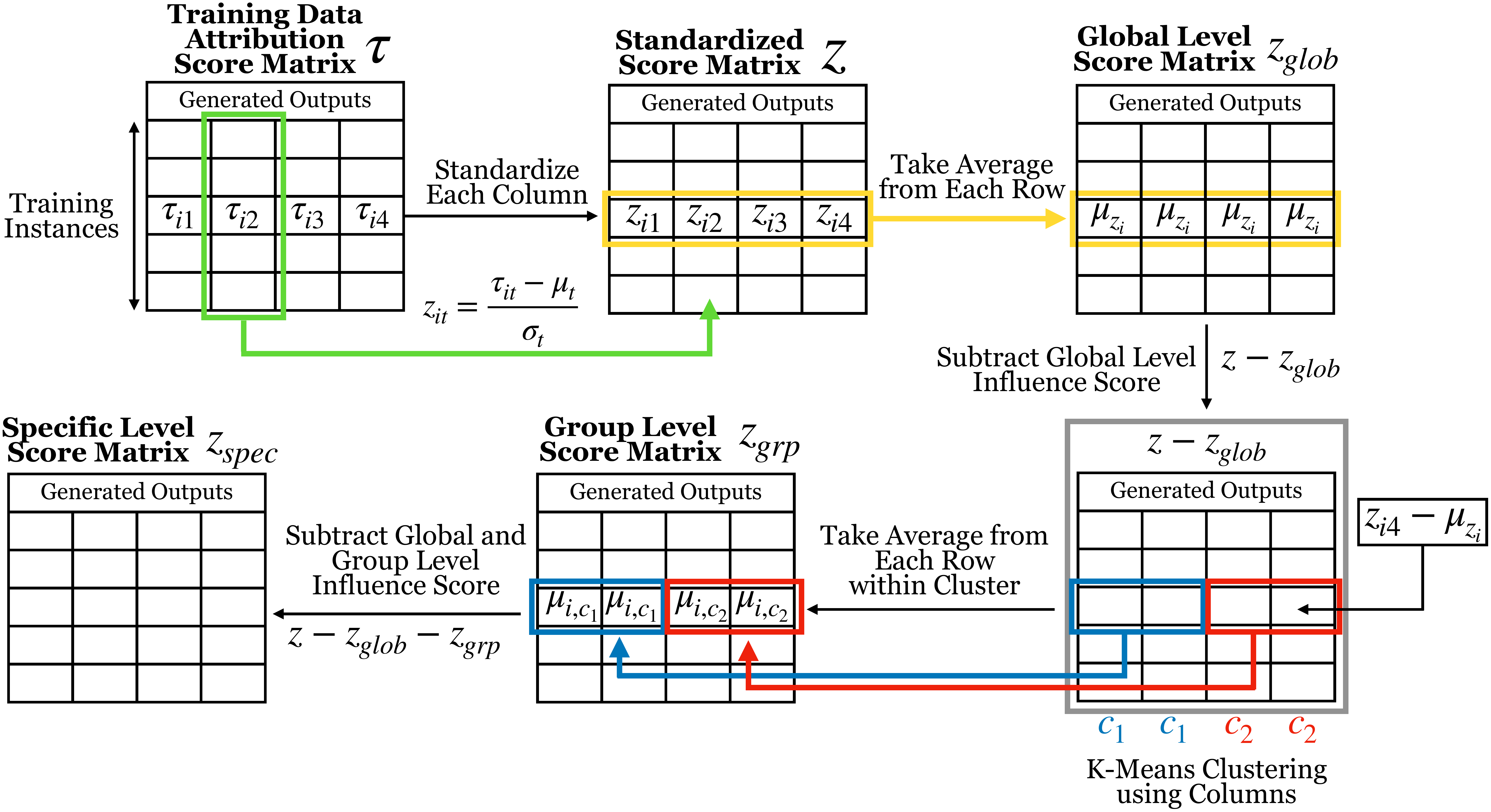}
  \caption{Decomposition of the standardized score matrix into the three levels.}
  \label{fig:decomposition}
\end{figure}

\subsubsection{Global Influence in Training Data Attribution}
\label{sec:global-influence}

Several studies on TDA have reported that a certain subset of the training instances tends to keep appearing as influential across almost every generated output, referred to as global influence~\cite{barshan2020relatif,hammoudeh2022identifying,hammoudeh2024survey}.
This global influence reflects the model's global behavior rather than any one generated output~\cite{barshan2020relatif,hammoudeh2022identifying}.
In terms of functional dependency, global influence corresponds to the dependency of the model itself on these training instances.
Gradient-based TDA methods usually include dot products of gradients in their influence computation, and prior work attributes global influence to the large gradient norms of certain training instances, which lets their global influence dominate the TDA scores~\cite{hanawa2021evaluation,hammoudeh2022identifying,hammoudeh2024survey}.
Previous approaches focus on identifying and mitigating the effect of globally influential training instances to obtain more individualized TDA scores for specific generated outputs.
Some replace the dot product of gradients with their cosine similarity or apply normalization to gradients, which mitigates the effect of the gradient norm~\cite{hanawa2021evaluation,hammoudeh2022identifying,akyurek-etal-2022-towards,xia2024less,barshan2020relatif,choe2024logra}.
Other methods find the gradient vector directions that many generated outputs have in common and use them to re-weight scores~\cite{chang2025scalable}.
These findings suggest that TDA scores may mix different levels of dependency, such as the dependency of the model itself and that of a specific generated output. Using TDA scores as a basis for remuneration therefore requires decomposing them into different levels of influence, each representing one scope of dependency. This supports transparency about which dependency the remuneration reflects.

\section{Decomposing the Training Data Attribution Scores into Levels of Influence}
\label{sec:decomposition}

We propose a decomposition of the standardized TDA score matrix into three levels, which are \Global, \Group and \Specific.
This decomposition allows us to analyze the influence at different levels in the TDA scores, revealing the hierarchical nature of influence that is already mixed in each score.
Each level of influence is related to a different scope of generated outputs.
The \Global level captures the overall influence of each training instance across all generated outputs, averaging training instance scores over the entire set of generated outputs.
The \Group level captures the influence within clusters of generated outputs, by clustering the score vectors of the generated outputs after the \Global level is removed.
The \Specific level captures the unique influence of each training instance on each individual generated output, after removing the \Global-level and \Group-level influence.
\Cref{fig:decomposition} illustrates the decomposition.

We consider $N$ training instances, indexed by $i$, and $T$ generated outputs, indexed by $t$.
TDA gives each pair of a training instance $i$ and a generated output $t$, a score $\tau_{it}$.
Since raw TDA scores differ in mean and scale across generated outputs, we standardize them before computing the three levels.
Each score is therefore standardized over the $N$ training instances,
\begin{equation}
  z_{it} = \frac{\tau_{it} - \mu_t}{\sigma_t},
  \label{eq:standardize}
\end{equation}
where $\mu_t$ and $\sigma_t$ are the mean and standard deviation of the scores for generated output $t$.
Standardization removes these differences but keeps the ranks within each generated output.
The standardized scores form the score matrix $\mathbf{Z} \in \mathbb{R}^{N \times T}$, whose entry for training instance $i$ and generated output $t$ is $z_{it}$.

We decompose each standardized score into a \Global, a \Group and a \Specific level score,
\begin{equation}
  z_{it} = z^{\mathrm{glob}}_{i} + z^{\mathrm{grp}}_{i,c(t)} + z^{\mathrm{spec}}_{it},
  \label{eq:split}
\end{equation}
where $c(t)$ is the cluster of generated output $t$. The \Global level score of the training instance $i$ is the mean of its scores over all $T$ generated outputs,
\begin{equation}
  z^{\mathrm{glob}}_{i} = \frac{1}{T} \sum_{t=1}^{T} z_{it}.
  \label{eq:glob}
\end{equation}
Each generated output $t$ is represented by the vector of its scores over the $N$ training instances after the \Global level is removed, $\mathbf{z}_t - \mathbf{z}^{\mathrm{glob}} = (z_{1t} - z^{\mathrm{glob}}_{1}, \dots, z_{Nt} - z^{\mathrm{glob}}_{N}) \in \mathbb{R}^{N}$.
We cluster these $T$ vectors with $k$-means and refer to these clusters as TDA score clusters.
The \Group level score of training instance $i$ for cluster $c$ is the mean of $z_{it} - z^{\mathrm{glob}}_{i}$ over the generated outputs in that cluster,
\begin{equation}
  z^{\mathrm{grp}}_{i,c} = \frac{1}{|G_c|} \sum_{t \in G_c} \bigl( z_{it} - z^{\mathrm{glob}}_{i} \bigr),
  \label{eq:grp}
\end{equation}
where $G_c$ is the set of generated outputs in cluster $c$.
The \Global level is subtracted first so that it is not counted again within the \Group level.
Since the same value $z^{\mathrm{glob}}_{i}$ is subtracted from every score of training instance $i$, the differences between its scores for different generated outputs do not change.
The \Specific level score is the rest,
\begin{equation}
  z^{\mathrm{spec}}_{it} = z_{it} - z^{\mathrm{glob}}_{i} - z^{\mathrm{grp}}_{i,c(t)}.
  \label{eq:spec}
\end{equation}
Within a TDA score cluster, the \Specific level keeps the differences between the scores of the generated outputs, because the same $z^{\mathrm{grp}}_{i,c}$ is subtracted from all of them.

The decomposition also shows how much of the scores each level holds over the whole score matrix.
We call this amount the proportion of the level.
The proportion uses squares to measure the size of each level relative to the total variance of the scores.
\begin{equation}
  s^{\mathrm{glob}} = \frac{\sum_{i,t} \bigl(z^{\mathrm{glob}}_{i}\bigr)^{2}}{\sum_{i,t} z_{it}^{2}}, \qquad
  s^{\mathrm{grp}} = \frac{\sum_{i,t} \bigl(z^{\mathrm{grp}}_{i,c(t)}\bigr)^{2}}{\sum_{i,t} z_{it}^{2}}, \qquad
  s^{\mathrm{spec}} = \frac{\sum_{i,t} \bigl(z^{\mathrm{spec}}_{it}\bigr)^{2}}{\sum_{i,t} z_{it}^{2}},
  \label{eq:size}
\end{equation}
where each sum runs over all $N$ training instances and all $T$ generated outputs.
Because each level is computed after the previous levels are subtracted from the scores, the three proportions add up to one, $s^{\mathrm{glob}} + s^{\mathrm{grp}} + s^{\mathrm{spec}} = 1$.

\section{Experiments}
\label{sec:setup}

\subsection{Experimental Setup}
\label{sec:setup-data}

\begin{table}[t]
  \footnotesize
  \caption{Experimental setup.}
  \label{tab:setup}
  \centering
  \begin{tabularx}{\linewidth}{>{\raggedright\arraybackslash}p{2.6cm}>{\raggedright\arraybackslash}X>{\raggedright\arraybackslash}X>{\raggedright\arraybackslash}X>{\raggedright\arraybackslash}X}
    \toprule[1.2pt]
    Model & \multicolumn{2}{>{\hsize=\dimexpr2\hsize+2\tabcolsep\relax\raggedright\arraybackslash}X}{MusicGen~\cite{copet2023musicgen}} & \multicolumn{2}{>{\hsize=\dimexpr2\hsize+2\tabcolsep\relax\raggedright\arraybackslash}X}{InspireMusic~\cite{zhang2025inspiremusic}} \\
    \cmidrule(l){2-3}\cmidrule(l){4-5}
     & MusicGen-S & MusicGen-L & InspireMusic-Base & InspireMusic-1.5B \\
    \midrule
    Fine-tuned Parameters & 0.4B & 3.3B & 0.5B & 1.5B \\
    \midrule
    Frozen Components & \multicolumn{2}{>{\hsize=\dimexpr2\hsize+2\tabcolsep\relax\raggedright\arraybackslash}X}{T5 text encoder~\cite{raffel2020exploring}, EnCodec~\cite{defossez2022high}} & \multicolumn{2}{>{\hsize=\dimexpr2\hsize+2\tabcolsep\relax\raggedright\arraybackslash}X}{WavTokenizer~\cite{ji2025wavtokenizer}, SRFM~\cite{zhang2025inspiremusic}} \\
    \midrule
    Training Data & \multicolumn{4}{>{\hsize=\dimexpr4\hsize+6\tabcolsep\relax\raggedright\arraybackslash}X}{30,000 ten-second segments from 10,000 MTG-Jamendo tracks~\cite{bogdanov2019mtgjamendo}} \\
    \midrule
    Generation & \multicolumn{4}{>{\hsize=\dimexpr4\hsize+6\tabcolsep\relax\raggedright\arraybackslash}X}{$T = 480$, from 8 prompts for each of 12 genres with $m = 5$ random seeds} \\
    \midrule
    TDA Methods & \multicolumn{4}{>{\hsize=\dimexpr4\hsize+6\tabcolsep\relax\raggedright\arraybackslash}X}{GradDot~\cite{charpiat2019input}, \GradCos~\cite{hanawa2021evaluation}, TracIn~\cite{pruthi2020estimating}, \LoGra~\cite{choe2024logra}, \FactGraSS~\cite{hu2026grass}, \Trak~\cite{park2023trak} ($\lambda = 1$), and the \Trak variants $\lambda = 0.1$, $\lambda = 0.01$ and ensemble} \\
    \midrule
        Deletion Test & \multicolumn{4}{>{\hsize=\dimexpr4\hsize+6\tabcolsep\relax\raggedright\arraybackslash}X}{25 deletion sets of 3,000 training instances (10\%) each, for GradDot and \Trak ($\lambda = 1$)} \\
    \bottomrule[1.2pt]
  \end{tabularx}
\end{table}

\subsubsection{Models and Training Data}
\begin{figure}[t]
  \centering
  \includegraphics[width=\linewidth]{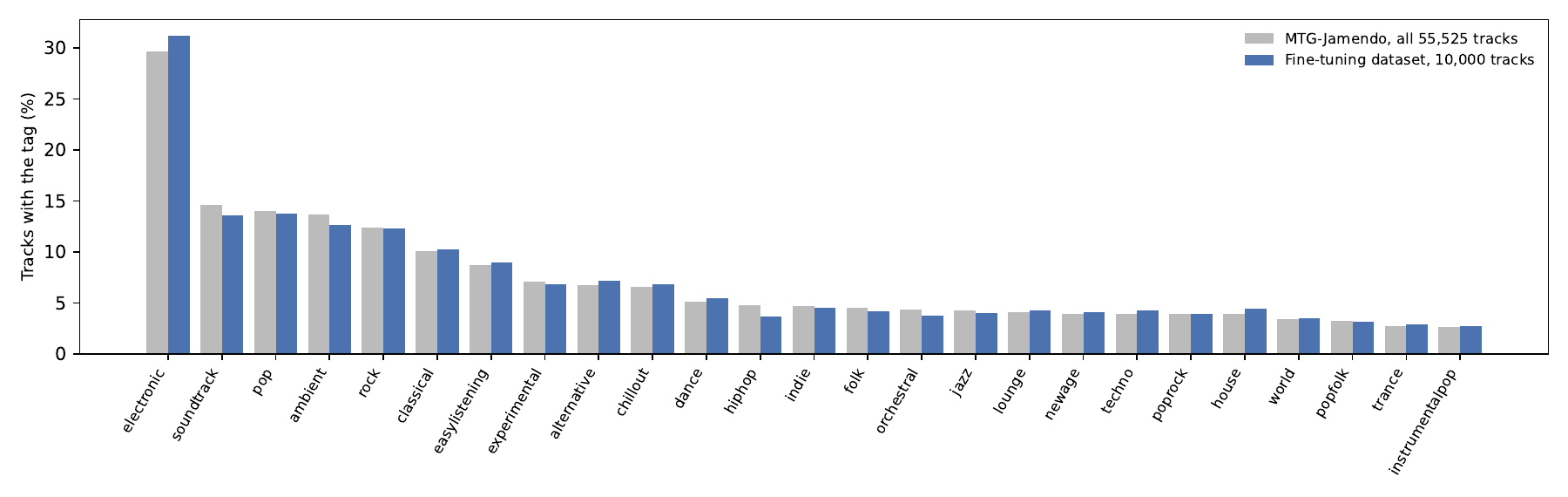}
  \caption{Distribution of genre tags in the whole MTG-Jamendo dataset (55,525 tracks) and in the fine-tuning dataset (10,000 tracks), for the 25 most frequent genre tags. A track can have several genre tags, so the percentages do not sum to 100.}
  \label{fig:catalogue}
\end{figure}

We use two music generative models, MusicGen~\cite{copet2023musicgen} and InspireMusic~\cite{zhang2025inspiremusic}, each in two sizes, and \Cref{tab:setup} summarizes the experimental setup.
Both generate audio tokens autoregressively, which are then decoded into audio waveforms by their respective audio token decoders.
MusicGen encodes the prompt with a T5 text encoder~\cite{raffel2020exploring}, and its transformer predicts EnCodec tokens~\cite{defossez2022high} conditioned on this encoding.
InspireMusic has no separate text encoder, and its transformer takes the text tokens of the prompt directly as input and predicts WavTokenizer tokens~\cite{ji2025wavtokenizer}, which the Super-Resolution Flow-Matching (SRFM) model~\cite{zhang2025inspiremusic} renders as audio.
Fine-tuning updates all parameters of the transformer in each model. The parameters of the other components are kept frozen and are not attributed. We refer to each ten-second segment of the training data as one training instance.
\Cref{fig:catalogue} shows that the distribution of genre tags in our fine-tuning dataset of 10,000 sampled tracks is similar to that of the whole MTG-Jamendo dataset.
We fine-tune each model on this training set with three random seeds.
The fine-tuning hyperparameters are reported in Appendix~\ref{app:finetuning}.

In this work, we focus on autoregressive music generative models.
An autoregressive model computes the log-likelihood of a generated output given its prompt exactly, as a sum over its tokens.
The gradient of the negative log-likelihood therefore defines the output gradient without changing over timesteps.
A diffusion-based model generates an output over iterative denoising steps which lead to the attribution of its generated output changing with the timestep~\cite{georgiev2023journey,xie2024data}.
Such models therefore need attribution methods adapted to the denoising process, and we leave them to future work. After fine-tuning, each model generates $T = 480$ outputs from text prompts.
The prompts are tag strings of MTG-Jamendo tracks not used in fine-tuning, and the first tag of each prompt gives its genre.
For example, the prompt ``folk, singersongwriter. instruments: classicalguitar.'' has the genre folk.
We draw 8 prompts for each of 12 genres and generate one output from each prompt with each of $m = 5$ random seeds.

\subsubsection{TDA Methods}
We then compute the TDA score of every training instance for every generated output with the six existing TDA methods and three variants of \Trak in \Cref{tab:setup}.
Each method computes the raw TDA score $\tau_{it}$ from the fine-tuned model. GradDot computes the TDA score of a training instance by the dot product of the two gradients which are the training gradient of instance $i$ and the output gradient of generated output $t$~\cite{charpiat2019input}.
The training gradient is the gradient of the training loss on instance $i$, and the output gradient is the gradient of the negative log-likelihood of generated output $t$ given its prompt.
\GradCos uses their cosine instead, which keeps only the angle between the two gradients and removes the effect of their norms~\cite{hanawa2021evaluation}.
TracIn sums the scores computed in the same way as GradDot at several checkpoints~\cite{pruthi2020estimating}.
This allows tracing the influence of a training instance over the training process rather than only at its end, and we use the four checkpoints saved during fine-tuning.
\Trak~\cite{park2023trak} add a weighting to the dot product of GradDot with random projections of all the training gradients and a constant $\lambda$ that sets how strongly it applies.
We use $\lambda = 1$ for \Trak, and $\lambda = 0.1$ and $0.01$ give two more variants, since a smaller $\lambda$ makes the weighting stronger.
\Trak can also average scores over several trained models~\cite{park2023trak}. We refer to this as the \Trak ensemble, which averages the $\lambda = 1$ scores of the three fine-tuned models.

Influence functions~\cite{koh2017understanding} start instead from how the trained parameters would change if a training instance were weighted slightly more, and estimate how this change affects the negative log-likelihood of a generated output.
This parameter change can be approximated by the training gradient multiplied by the inverse of the Hessian of the training loss.
The Hessian is a matrix whose size is the square of the number of model parameters, which makes its direct computation infeasible for large models.
\LoGra~\cite{choe2024logra} and \FactGraSS~\cite{hu2026grass} approximate it by the covariance of the training gradients within each layer.
The two methods differ in how they compress the gradients to make this computation feasible.
All scores except TracIn and the \Trak ensemble come from the final checkpoint of one fine-tuned model. The hyperparameters of the TDA methods are reported in Appendix~\ref{app:tda}.

\subsubsection{Deletion Test}
The deletion test removes the training instances that score highest in one level and fine-tunes the pretrained model again without them~\cite{chen2026mechanistic,kowal2026concept}.
It then measures the log-likelihood (LL) of the generated outputs that the level should affect, which we call its target generated outputs.
The LL of a generated output is its mean per-token log-likelihood under the retrained model.
As a reference, we also retrain the model without a random set of the same size, with the same retraining seed.
For each generated output $t$, we compute
\begin{equation*}
  \Delta\mathrm{LL}_t = \mathrm{LL}_t(\text{without the deletion set}) - \mathrm{LL}_t(\text{without the random set}).
\end{equation*}
If the level has an effect on its target generated outputs, we expect $\Delta\mathrm{LL}_t$ to be negative for them.

The \Global level deletion set targets all generated outputs and tests whether their mean $\Delta\mathrm{LL}$ is negative.
A drop shared by all generated outputs, however, does not show an effect particular to one TDA score cluster or one prompt.
The \Group- and \Specific-level deletion sets are therefore compared with the other generated outputs within the same scope.
Write $\overline{\Delta\mathrm{LL}}(\cdot)$ for the mean $\Delta\mathrm{LL}$ over a set of generated outputs and $G_c$ for the generated outputs in TDA score cluster $c$ as above.
Write $P_p$ for the $m$ generated outputs of a targeted prompt $p$, one for each random seed.
The three levels are then measured by
\begin{equation}
  \begin{gathered}
    D^{\mathrm{glob}} = \frac{1}{T} \sum_{t=1}^{T} \Delta\mathrm{LL}_t,
    \\
    D^{\mathrm{grp}}_{c} = \overline{\Delta\mathrm{LL}}(G_c) - \frac{1}{k-1} \sum_{c' \neq c} \overline{\Delta\mathrm{LL}}(G_{c'}), \qquad
    D^{\mathrm{spec}}_{p} = \frac{1}{m} \sum_{t \in P_p} \Bigl( \Delta\mathrm{LL}_t - \overline{\Delta\mathrm{LL}}(Q_t) \Bigr),
  \end{gathered}
  \label{eq:deletion}
\end{equation}
where $k$ is the number of TDA score clusters and $Q_t$ is the set of generated outputs of the other prompts in TDA score cluster $c(t)$.
We set the number of TDA score clusters to $k = 12$, the same number as the genres, so that the TDA score clusters can be compared with genre labels.
Comparing $t$ only with the other generated outputs in $c(t)$ separates the drop particular to the targeted prompt from the drop shared by all generated outputs in $c(t)$.
A level with an effect gives a negative $D^{\mathrm{glob}}$, $D^{\mathrm{grp}}_{c}$ or $D^{\mathrm{spec}}_{p}$.
The \Global level has one deletion set, while the \Group and \Specific levels have one deletion set for each TDA score cluster and each targeted prompt.
The 12 targeted prompts are drawn at random from the 96 prompts and are the same for all models and methods.
We report the \Group and \Specific levels by their means over the deletion sets,
\begin{equation}
  \bar{D}^{\mathrm{grp}} = \frac{1}{k} \sum_{c=1}^{k} D^{\mathrm{grp}}_{c}, \qquad
  \bar{D}^{\mathrm{spec}} = \frac{1}{12} \sum_{j=1}^{12} D^{\mathrm{spec}}_{p_j},
  \label{eq:deletion-mean}
\end{equation}
where $p_1, \dots, p_{12}$ are the 12 targeted prompts.

\Cref{tab:setup} summarizes the deletion sets of each method.
Training instances are ranked by $z^{\mathrm{glob}}_{i}$ for the \Global set, by $z^{\mathrm{grp}}_{i,c}$ for the \Group set of TDA score cluster $c$, and by $z^{\mathrm{spec}}_{it}$ averaged over the $m$ generated outputs of a targeted prompt for its \Specific set.
The model is retrained with three seeds for MusicGen-S, MusicGen-L and InspireMusic.
Each deletion set needs its own retraining, so we select two of the six methods and three variants, GradDot and \Trak with $\lambda = 1$.
In Table~\ref{tab:proportions}, GradDot is among the methods with the highest \Group proportions and the \Trak variants have the lowest.
The GPU time of fine-tuning and of the deletion test is reported in Appendix~\ref{app:compute}.

\subsection{Global Level Analysis}

To investigate where high \Global level scores come from, we look at the norms of the training gradients and their alignment with the common direction of the output gradients.
Gradient-based TDA methods score a training instance by an inner product of its training gradient and the output gradient.
Previous studies link the high scores of globally influential training instances to their large gradient norm~\cite{barshan2020relatif,hanawa2021evaluation,hammoudeh2022identifying}.
However, the inner product of two vectors is the product of their norms and the cosine similarity between them.
A high score can therefore come from a large norm of the training gradient or from a high cosine similarity with the output gradient. Let $\mathbf{g}_i$ be the training gradient of training instance $i$ and $\mathbf{g}_t$ the output gradient of generated output $t$.
The gradient norm of training instance $i$ is $\lVert \mathbf{g}_i \rVert$.
The alignment requires the common direction of the output gradients, it can be captured by their mean direction.
To prevent output gradients with large norms from dominating the mean, we divide each output gradient by its norm before averaging them.
We measure the alignment of training instance $i$ by the cosine similarity between its training gradient and this direction,
\begin{equation}
  \cos(\mathbf{g}_i, \mathbf{d}) = \frac{\mathbf{g}_i^{\top} \mathbf{d}}{\lVert \mathbf{g}_i \rVert \, \lVert \mathbf{d} \rVert},
  \qquad
  \mathbf{d} = \frac{1}{T} \sum_{t=1}^{T} \frac{\mathbf{g}_t}{\lVert \mathbf{g}_t \rVert},
  \label{eq:alignment}
\end{equation}
where $\mathbf{d}$ is the common direction.

For each method and model, we compare $z^{\mathrm{glob}}_{i}$ with the gradient norm and the alignment using the Spearman rank correlation over the $N$ training instances,
\begin{equation}
  \rho^{\mathrm{norm}} = \mathrm{Spearman}\bigl( z^{\mathrm{glob}}_{i}, \lVert \mathbf{g}_i \rVert \bigr),
  \qquad
  \rho^{\mathrm{align}} = \mathrm{Spearman}\bigl( z^{\mathrm{glob}}_{i}, \cos(\mathbf{g}_i, \mathbf{d}) \bigr).
  \label{eq:spearman}
\end{equation}
Each correlation is 1 when the two rankings are identical, 0 when they are unrelated and $-1$ when one is the reverse of the other.

\subsection{Group Level Analysis}
\label{sec:setup-size}

To see whether the TDA score clusters formed by $k$-means clustering relate to existing features of the generated outputs, we compare them with genre and other audio characteristics.
The clusters are formed by treating the scores $\mathbf{z}_t - \mathbf{z}^{\mathrm{glob}}$ of the training instances for each generated output as a vector, and generated outputs are assigned to the clusters.
Genre and each audio feature also divide the same $T = 480$ generated outputs into clusters.
We measure the alignment between the clusters from different criteria to ask whether they place the same generated outputs in the same cluster, using the Wallace coefficient~\cite{wallace1983comment}\footnote{The Wallace coefficient is the pairwise precision of one clustering with respect to another, computed here with scikit-learn~\cite{pedregosa2011scikit}.}.

To compare the TDA score clusters with an existing audio feature, we first extract audio features from each generated output.
The extracted audio features are then used to form a clustering of the generated outputs.
We use three audio features, the MFCCs~\cite{davis1980mfcc}, the log-mel spectrogram and the chroma features~\cite{muller2015fundamentals}.
Each feature is clustered with $k$-means into $k = 12$ feature clusters, the same number as the TDA score clusters.
Each genre also forms one feature cluster.
The details of the clustering and the audio features are reported in Appendix~\ref{app:features}.
The Wallace coefficient compares the TDA score clusters with the feature clusters through pairs of generated outputs.
Write $c(t)$ for the TDA score cluster of generated output $t$ and $f(t)$ for its feature cluster.
We compute $W$ separately for genre and for each audio feature.
We count the pairs of generated outputs in the same TDA score cluster, $n_{\mathrm{score}}$, and the pairs among them that are also in the same feature cluster, $n_{\mathrm{both}}$,
\begin{equation}
  W = \frac{n_{\mathrm{both}}}{n_{\mathrm{score}}},
  \qquad
  n_{\mathrm{score}} = \sum_{t \neq t'} \mathbf{1}\bigl[c(t) = c(t')\bigr],
  \qquad
  n_{\mathrm{both}} = \sum_{t \neq t'} \mathbf{1}\bigl[c(t) = c(t')\bigr] \, \mathbf{1}\bigl[f(t) = f(t')\bigr],
  \label{eq:wallace}
\end{equation}
where $\mathbf{1}[\cdot]$ is 1 if the condition holds and 0 otherwise.
$W$ is therefore the probability that two generated outputs in the same TDA score cluster are also in the same feature cluster.
A value of $W = 1$ means that every pair in a TDA score cluster is also in the same feature cluster.

\section{Results}
\label{sec:results}

\subsection{Proportion of Each Level}

\begin{table}[t]
  \footnotesize
  \caption{Proportion of the \Global and \Group levels (\Group by TDA score clusters, $k = 12$). S, L, IM and IM-1.5B denote MusicGen-S, MusicGen-L, InspireMusic-Base and InspireMusic-1.5B.}
  \label{tab:proportions}
  \centering
  \begin{tabular}{lcccccccc}
    \toprule
     & \multicolumn{4}{c}{\Global} & \multicolumn{4}{c}{\Group} \\
    \cmidrule(lr){2-5} \cmidrule(lr){6-9}
    Method & S & L & IM & IM-1.5B & S & L & IM & IM-1.5B \\
    \midrule
    Random & 0.002 & 0.002 & 0.002 & 0.002 & 0.02 & 0.02 & 0.02 & 0.02 \\
    GradDot & 0.106 & 0.131 & 0.360 & 0.403 & 0.212 & 0.151 & 0.113 & 0.118 \\
    \GradCos & 0.106 & 0.128 & 0.359 & 0.401 & 0.209 & 0.148 & 0.113 & 0.119 \\
    TracIn & \textbf{0.135} & \textbf{0.139} & \textbf{0.454} & \textbf{0.651} & \textbf{0.239} & \textbf{0.214} & \textbf{0.128} & \textbf{0.173} \\
    \LoGra & 0.059 & 0.070 & 0.132 & 0.134 & 0.073 & 0.070 & 0.076 & 0.101 \\
    \FactGraSS & 0.054 & 0.058 & 0.114 & 0.115 & 0.063 & 0.055 & 0.075 & 0.089 \\
    \Trak $\lambda=1$ & 0.032 & 0.023 & 0.042 & 0.036 & 0.038 & 0.032 & 0.041 & 0.048 \\
    \Trak $\lambda=0.1$ & 0.021 & 0.015 & 0.022 & 0.023 & 0.034 & 0.029 & 0.036 & 0.043 \\
    \Trak $\lambda=0.01$ & 0.019 & 0.014 & 0.019 & 0.022 & 0.033 & 0.028 & 0.035 & 0.041 \\
    \Trak (ensemble) & 0.033 & 0.023 & 0.043 & 0.040 & 0.040 & 0.032 & 0.040 & 0.052 \\
    \bottomrule
  \end{tabular}
\end{table}

\Cref{tab:proportions} reports the \Global and \Group proportions of the six methods and three variants in each model, with the random baselines in the row Random.
Random baselines provide a reference point to assess whether the observed proportions exceed what would be expected by chance.
The \Global level baseline is $1/T \approx 0.002$, which is the \Global proportion that the mean over $T$ generated outputs yields by itself, regardless of the model behavior or the TDA method.
The \Group level baseline is $(k-1)/(T-1) \times (1 - s^{\mathrm{glob}}) \approx 0.02$, which is the expected proportion when the generated outputs are assigned randomly to $k = 12$ groups of the same sizes.

Our results show that the \Global level and the \Group level are observed consistently across all models and methods, with proportions above their random baselines.
However, their relative proportions vary across methods.
Within each model, both proportions are largest for GradDot, \GradCos and TracIn, which compute the TDA scores from the unweighted dot product of gradients between the training instances and the generated outputs.
The methods that weight this dot product, namely \LoGra, \FactGraSS and the \Trak variants, have smaller \Global and \Group proportions.
A proportion describes how the scores are composed of the three levels, but it does not directly indicate how strongly a level influences the model's behavior on the generated outputs in its scope.
Whether each level changes the model's behavior on those generated outputs is examined through the deletion test.

\subsection{Deletion Test Results}

\begin{table}[t]
  \footnotesize
  \caption{Deletion test results ($\times 10^{3}$). \Trak uses $\lambda = 1$.}
  \label{tab:deletion}
  \centering
  \begin{tabular}{lcccccccc}
    \toprule
     & \multicolumn{2}{c}{MusicGen-S} & \multicolumn{2}{c}{MusicGen-L} & \multicolumn{2}{c}{InspireMusic-Base} & \multicolumn{2}{c}{InspireMusic-1.5B} \\
    \cmidrule(lr){2-3} \cmidrule(lr){4-5} \cmidrule(lr){6-7} \cmidrule(lr){8-9}
    Level & GradDot & \Trak & GradDot & \Trak & GradDot & \Trak & GradDot & \Trak \\
    \midrule
    \Global, $D^{\mathrm{glob}}$ & $-1.79$ & $-3.42$ & $-0.88$ & $-2.42$ & $-14.86$ & $-16.79$ & $-11.66$ & $-18.84$ \\
    \Group, $\bar{D}^{\mathrm{grp}}$ & $-1.15$ & $-3.13$ & $-0.68$ & $-2.48$ & $-6.14$ & $-10.48$ & $-7.46$ & $-9.98$ \\
    \Specific, $\bar{D}^{\mathrm{spec}}$ & $-5.89$ & $-7.23$ & $-2.81$ & $-3.21$ & $-22.24$ & $-23.12$ & $-9.83$ & $-11.80$ \\
    \bottomrule
  \end{tabular}
\end{table}

\begin{figure}[h]
  \centering
  \includegraphics[width=\linewidth]{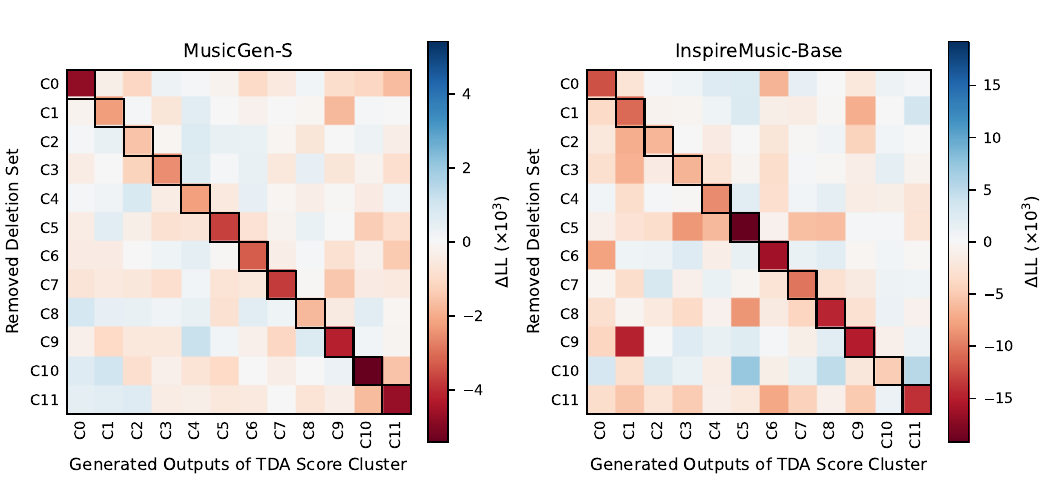}
  \caption{Results of the \textit{group}-level deletion test for \Trak with $\lambda = 1$ in MusicGen-S (left) and InspireMusic-Base (right). Each row is the removed deletion set of a TDA score cluster, and each column is the generated outputs of a TDA score cluster. Each entry is the mean $\Delta\mathrm{LL}$ ($\times 10^{3}$) over the generated outputs of the column and three retraining seeds. Outlined cells are the diagonal.}
  \label{fig:heatmap}
\end{figure}

We test whether each level of influence holds a counterfactual effect on the model's behavior in its scope through the deletion test~\cite{chen2026mechanistic,kowal2026concept}. Table~\ref{tab:deletion} reports $D^{\mathrm{glob}}$, $\bar{D}^{\mathrm{grp}}$ and $\bar{D}^{\mathrm{spec}}$ (\Cref{eq:deletion,eq:deletion-mean}), averaged over the retraining seeds.
All entries are negative, and every entry with three seeds is negative in each seed.
A negative value means that removing the deletion set lowers the LL of the generated outputs within the scope of the level more than the reference of the level.
For the \Global level, the value is the LL of all the generated outputs under the model fine-tuned without the deletion set minus that under the model fine-tuned without the random set.
At the \Group level, the value is this difference for the generated outputs of the targeted TDA score cluster minus that for the generated outputs of the other TDA score clusters.
At the \Specific level, the value is this difference for the generated outputs of the targeted prompt minus that for the generated outputs of the other prompts in the same TDA score cluster.
In MusicGen-S, for example, the \Global, \Group and \Specific entries are $-1.79$, $-1.15$ and $-5.89$ for GradDot, and $-3.42$, $-3.13$ and $-7.23$ for \Trak.
With these results, we observe that each level of influence has a counterfactual effect on the model's behavior within its scope.

\Cref{fig:heatmap} shows deletion test results for \Trak in MusicGen-S and InspireMusic-Base in every deletion set separately.
If the \Group level has an effect within its scope, the largest drop in each row should appear on the diagonal, which is the targeted TDA score cluster.
The diagonal entry is the lowest in its row in all 12 rows for MusicGen-S and in 10 of the 12 rows for InspireMusic-Base.
Removing the deletion set of a TDA score cluster thus lowers the LL of the generated outputs of that cluster more than those of the other TDA score clusters.

\subsection{Global Level Alignment}

\begin{table}[t]
  \footnotesize
  \caption{Norm of the common direction $\lVert \mathbf{d} \rVert$, and Spearman correlations $\rho^{\mathrm{align}}$ and $\rho^{\mathrm{norm}}$ of $z^{\mathrm{glob}}_{i}$.}
  \label{tab:alignment}
  \centering
  \begin{tabular}{lcccccccc}
    \toprule
     & \multicolumn{2}{c}{MusicGen-S} & \multicolumn{2}{c}{MusicGen-L} & \multicolumn{2}{c}{InspireMusic-Base} & \multicolumn{2}{c}{InspireMusic-1.5B} \\
    \cmidrule(lr){2-3} \cmidrule(lr){4-5} \cmidrule(lr){6-7} \cmidrule(lr){8-9}
    $\lVert \mathbf{d} \rVert$ & \multicolumn{2}{c}{0.208} & \multicolumn{2}{c}{0.192} & \multicolumn{2}{c}{0.290} & \multicolumn{2}{c}{0.282} \\
    \midrule
    Method & $\rho^{\mathrm{align}}$ & $\rho^{\mathrm{norm}}$ & $\rho^{\mathrm{align}}$ & $\rho^{\mathrm{norm}}$ & $\rho^{\mathrm{align}}$ & $\rho^{\mathrm{norm}}$ & $\rho^{\mathrm{align}}$ & $\rho^{\mathrm{norm}}$ \\
    \midrule
    GradDot & \textbf{1.00} & $-0.06$ & \textbf{1.00} & $-0.06$ & \textbf{1.00} & $+0.14$ & 0.99 & $+0.15$ \\
    \GradCos & \textbf{1.00} & $-0.06$ & \textbf{1.00} & $-0.05$ & \textbf{1.00} & $+0.14$ & \textbf{1.00} & $+0.14$ \\
    TracIn & 0.81 & $\mathbf{+0.04}$ & 0.70 & $\mathbf{+0.05}$ & 0.90 & $\mathbf{+0.32}$ & 0.85 & $\mathbf{+0.31}$ \\
    \LoGra & 0.61 & $-0.07$ & 0.46 & $-0.09$ & 0.71 & $+0.11$ & 0.44 & $+0.17$ \\
    \FactGraSS & 0.51 & $-0.07$ & 0.42 & $-0.09$ & 0.68 & $+0.11$ & 0.43 & $+0.18$ \\
    \Trak $\lambda=1$ & 0.58 & $-0.04$ & 0.62 & $-0.03$ & 0.60 & $+0.08$ & 0.52 & $+0.12$ \\
    \Trak $\lambda=0.1$ & 0.43 & $-0.02$ & 0.44 & $-0.02$ & 0.43 & $+0.06$ & 0.33 & $+0.07$ \\
    \Trak $\lambda=0.01$ & 0.39 & $-0.02$ & 0.40 & $-0.01$ & 0.38 & $+0.05$ & 0.29 & $+0.06$ \\
    \Trak (ensemble) & 0.57 & $-0.04$ & 0.60 & $-0.05$ & 0.58 & $+0.09$ & 0.51 & $+0.11$ \\
    \bottomrule
  \end{tabular}
\end{table}

We examine where the \Global level influence of training instances comes from.
Prior studies attribute globally influential training instances to their large gradient norms~\cite{barshan2020relatif,hanawa2021evaluation,hammoudeh2022identifying}.
A gradient-based score is an inner product of the training gradient and the output gradient, which is the product of their norms and the cosine similarity between them.
We therefore compare the \Global level score of each training instance with the norm of its training gradient and with its alignment, the cosine similarity between its training gradient and the common direction of the output gradients.
The common direction is the mean of the output gradients after each is divided by its norm.
This examination allows us to understand whether the \Global level influence is primarily due to the magnitude of the training gradients or their alignment with the common direction of the output gradients.
The comparison uses the Spearman rank correlation over the $N$ training instances, $\rho^{\mathrm{norm}}$ for the norm and $\rho^{\mathrm{align}}$ for the alignment.
Table~\ref{tab:alignment} reports $\rho^{\mathrm{align}}$ and $\rho^{\mathrm{norm}}$ of each method in each model, with the norm of the common direction $\lVert \mathbf{d} \rVert$ in the top row.

$\lVert \mathbf{d} \rVert$ is 0 when the output gradients share no direction and 1 when they all point in the same direction.
It is above 0 in every model, so the output gradients share a common direction.
This agrees with the shared gradient directions reported in prior work~\cite{yeh2022first,chang2025scalable,papyan2020traces}.
Our results show that the \Global level scores follow the alignment with this direction rather than the gradient norm, as $\rho^{\mathrm{align}}$ is consistently positive.
$\rho^{\mathrm{norm}}$ is close to 0, at most 0.18 in absolute value except for TracIn.
TracIn has $\rho^{\mathrm{norm}}$ of 0.32 and 0.31 in InspireMusic-Base and InspireMusic-1.5B, but its $\rho^{\mathrm{align}}$ in the same models is 0.90 and 0.85.
The $\rho^{\mathrm{align}}$ close to 1 for GradDot and \GradCos largely follows from their definitions, because their scores are computed directly from the output gradient.
Therefore, for these two methods, the interpretation of the results should primarily focus on the weak correlation $\rho^{\mathrm{norm}}$ rather than the strong correlation $\rho^{\mathrm{align}}$.

\subsection{Group Level Alignment}
\label{sec:clustering-alignment}

\begin{table}[t]
  \footnotesize
  \caption{Wallace coefficient $W$ between the TDA score clusters and the feature clusters of each criterion.}
  \label{tab:wallace}
  \centering
  \setlength{\tabcolsep}{3pt}
  \begin{tabular}{lcccccccccc}
    \toprule
     & & & & & & \Trak & \Trak & \Trak & \Trak & \\
    Model & GradDot & \GradCos & TracIn & \LoGra & \FactGraSS & $\lambda=1$ & $\lambda=0.1$ & $\lambda=0.01$ & (ensemble) & Random \\
    \midrule
    \multicolumn{11}{l}{\textit{Genre}} \\
    MusicGen-S & 0.103 & 0.104 & 0.118 & 0.166 & \textbf{0.201} & 0.149 & 0.141 & 0.139 & 0.143 & 0.081 \\
    MusicGen-L & 0.102 & 0.100 & 0.109 & 0.190 & \textbf{0.194} & 0.137 & 0.128 & 0.122 & 0.137 & 0.081 \\
    InspireMusic-Base & 0.102 & 0.116 & 0.107 & 0.163 & \textbf{0.179} & 0.175 & 0.161 & 0.161 & 0.168 & 0.081 \\
    InspireMusic-1.5B & 0.107 & 0.107 & 0.090 & \textbf{0.122} & 0.121 & 0.115 & 0.111 & 0.104 & 0.115 & 0.081 \\
    \midrule
    \multicolumn{11}{l}{\textit{MFCC}} \\
    MusicGen-S & 0.152 & 0.155 & 0.180 & 0.253 & \textbf{0.284} & 0.246 & 0.216 & 0.220 & 0.221 & 0.117 \\
    MusicGen-L & 0.142 & 0.140 & 0.147 & 0.195 & \textbf{0.209} & 0.176 & 0.176 & 0.167 & 0.184 & 0.122 \\
    InspireMusic-Base & 0.140 & 0.138 & 0.144 & 0.159 & 0.152 & \textbf{0.164} & 0.152 & 0.154 & 0.156 & 0.108 \\
    InspireMusic-1.5B & 0.176 & 0.179 & 0.147 & \textbf{0.215} & 0.207 & 0.182 & 0.180 & 0.177 & 0.189 & 0.119 \\
    \midrule
    \multicolumn{11}{l}{\textit{Log-mel spectrogram}} \\
    MusicGen-S & 0.147 & 0.145 & 0.159 & 0.199 & \textbf{0.211} & 0.187 & 0.172 & 0.166 & 0.174 & 0.099 \\
    MusicGen-L & 0.159 & 0.157 & 0.173 & \textbf{0.230} & 0.225 & 0.174 & 0.164 & 0.153 & 0.180 & 0.105 \\
    InspireMusic-Base & 0.128 & 0.132 & 0.142 & 0.156 & 0.149 & \textbf{0.157} & 0.142 & 0.146 & 0.150 & 0.102 \\
    InspireMusic-1.5B & 0.176 & 0.182 & 0.138 & \textbf{0.200} & 0.193 & 0.176 & 0.166 & 0.161 & 0.181 & 0.118 \\
    \midrule
    \multicolumn{11}{l}{\textit{Chroma}} \\
    MusicGen-S & 0.132 & 0.130 & 0.141 & 0.169 & \textbf{0.173} & 0.164 & 0.168 & 0.164 & 0.164 & 0.123 \\
    MusicGen-L & 0.116 & 0.114 & 0.124 & 0.149 & \textbf{0.151} & 0.139 & 0.146 & 0.133 & 0.144 & 0.107 \\
    InspireMusic-Base & 0.122 & 0.125 & 0.129 & \textbf{0.146} & 0.141 & 0.138 & 0.131 & 0.140 & 0.136 & 0.099 \\
    InspireMusic-1.5B & 0.143 & 0.146 & 0.134 & \textbf{0.167} & 0.162 & 0.157 & 0.144 & 0.151 & 0.152 & 0.122 \\
    \bottomrule
  \end{tabular}
\end{table}

\begin{figure}
  \centering
  \includegraphics[width=0.7\linewidth]{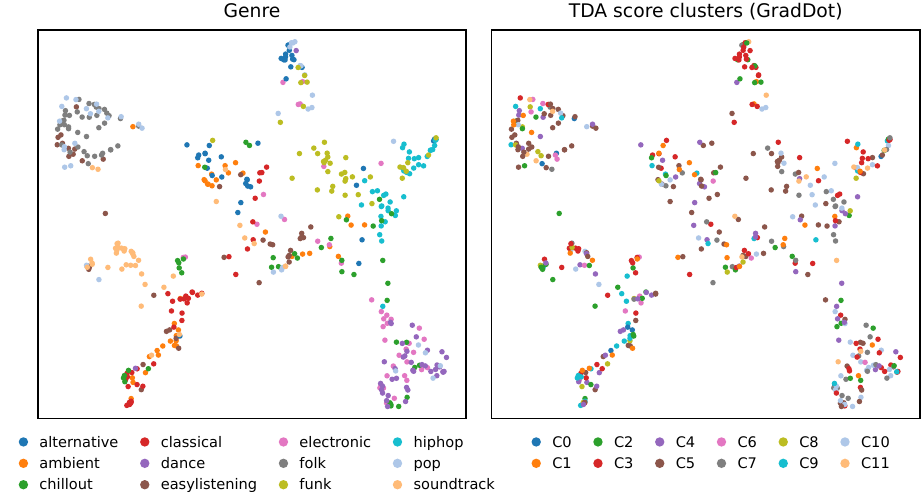}
  \caption{Generated outputs of MusicGen-S in the \CLAP embedding space with UMAP, coloured by genre (left) and by the TDA score clusters of GradDot (right). A colour of a TDA score cluster is only a cluster label and does not indicate a genre.}
  \label{fig:embed-clap-graddot}
\end{figure}

We examine whether the TDA score clusters relate to the feature clusters, which are formed from the genre of each prompt or from an audio feature of each generated output.
We compare the two with the Wallace coefficient $W$~\cite{wallace1983comment} (\Cref{eq:wallace}), the probability that two generated outputs in the same TDA score cluster are also in the same feature cluster.
Table~\ref{tab:wallace} reports $W$ of each method in each model for each clustering criterion, with the random baselines, the expected $W$ when the generated outputs are assigned randomly to the TDA score clusters of the same sizes, in the column Random.
Bold marks the largest $W$ in each row.
A value of $W$ close to 1 would mean that the TDA score clusters reproduce the feature clusters, while a value close to the random baseline means that the two group the generated outputs independently.
Our results show that $W$ is low in every model, method and criterion, and the largest value is 0.284.
$W$ also differs between methods in the same model.
In MusicGen-S, for example, $W$ for genre is 0.103 for GradDot, while \FactGraSS in MusicGen-S and MusicGen-L has the two largest values of $W$ for genre, 0.201 and 0.194.
It seems that \FactGraSS tends to have higher $W$ for genre, and this relation to genre partly appears in the audio features as well.
However, all values of $W$ are low, so the TDA score clusters are not related to genre or to the audio features.

\Cref{fig:embed-clap-graddot} depicts the generated outputs of MusicGen-S in the \CLAP~\cite{wu2023large} audio embedding space with UMAP~\cite{mcinnes2018umap}. The outputs are colored by genre and, in the same layout, by the TDA score clusters of GradDot.
Generated outputs of the same genre lie close to each other, while those of the same TDA score cluster are spread across the embedding space.
The TDA score clusters of GradDot therefore do not correspond to the genre of the generated outputs.
Together with the low $W$ in every model and method, this indicates that the TDA scores cluster the generated outputs by a criterion that the model itself forms, rather than by genre or audio characteristics.
This suggests that the model relates its generated outputs to the training data through functional relations beyond musical similarity.
More embedding visualizations are reported in Appendix~\ref{app:embedding}.

\subsection{Specific Level Analysis}
\label{sec:specific-change}

\begin{table}[t]
  \footnotesize
  \caption{Fraction of change in the 50 training instances with the highest \Specific level scores when the criterion changes from genre to the TDA score clusters.}
  \label{tab:specific-change}
  \centering
  \setlength{\tabcolsep}{3pt}
  \begin{tabular}{lccccccccc}
    \toprule
     & & & & & & \Trak & \Trak & \Trak & \Trak \\
    Model & GradDot & \GradCos & TracIn & \LoGra & \FactGraSS & $\lambda=1$ & $\lambda=0.1$ & $\lambda=0.01$ & (ensemble) \\
    \midrule
    MusicGen-S & 0.55 & 0.56 & 0.55 & 0.22 & 0.20 & 0.30 & 0.30 & 0.30 & 0.30 \\
    MusicGen-L & 0.50 & 0.52 & 0.54 & 0.22 & 0.20 & 0.29 & 0.30 & 0.30 & 0.29 \\
    InspireMusic-Base & 0.49 & 0.51 & 0.51 & 0.27 & 0.27 & 0.32 & 0.31 & 0.32 & 0.30 \\
    InspireMusic-1.5B & 0.53 & 0.57 & 0.51 & 0.36 & 0.35 & 0.33 & 0.33 & 0.33 & 0.34 \\
    \bottomrule
  \end{tabular}
\end{table}

Since the \Specific level is calculated as the residual after subtracting the \Group level, it depends on the criterion that defines the \Group level.
We measure how much the \Specific level changes when the criterion changes from genre to the TDA score clusters.
For each generated output, we take the 50 training instances with the highest \Specific level scores when genre defines the \Group level.
We measure the fraction of these 50 that are no longer among the highest 50 when the TDA score clusters define the \Group level, and average it over the generated outputs.
A value of 1 means that all 50 training instances change, and a value of 0 means that none of them change. Table~\ref{tab:specific-change} reports the fraction for each method in each model.
Our results show that about half of the 50 training instances change for GradDot, \GradCos and TracIn in every model, from 0.49 to 0.57.
Which training instances belong to the \Specific level of a generated output therefore depends on the criterion which defines how to form the \Group level.
A remuneration based on the \Specific level scores thus also depends on this \Group level criterion.
In MusicGen-S, for example, 0.55 of them change for GradDot.
For the other methods, from 0.20 to 0.36 of them change.

\section{Conclusion}
\label{sec:discussion}

We proposed a decomposition of TDA scores to support the analysis of what a remuneration based on each level would reflect.
Across four music generative models and six TDA methods and three variants, the scores contain a \Global and a \Group level.
In the deletion test, each level has an effect on the model's behavior within its scope.
This means that a single TDA score combines influence on the model itself, on a group of generated outputs and on a single generated output. The \Global level corresponds to the dependency of the model itself, which is closer in scope to a general payment for training on copyrighted works~\cite{geiger2024forgotten} than to a remuneration for any one specific generated output.
The \Group level is formed by the model and does not coincide with genre or simple audio features.
However, we do not identify what the TDA score clusters correspond to, and identifying what they represent remains future work.
The \Specific level changes with the criterion that defines the \Group level, so a remuneration based on it depends on that criterion.

Our work lays the groundwork for interpreting each level as contribution, which requires further discussion that involves legal, economic and ethical perspectives.
With our work, we aim to bridge the legal and socio-economic research on copyright and attribution with technical estimates of the influence of training data at each scope of dependency.
By separating the levels, our decomposition turns a single TDA score into distinct forms of influence that each of these discussions can examine.
This makes TDA a more transparent basis for remuneration in generative AI.

\section*{Ethical Considerations}
Our experiments involve no human subjects and no sensitive personal data.
All training data comes from the MTG-Jamendo dataset~\cite{bogdanov2019mtgjamendo}, whose audio files are licensed under Creative Commons licenses.
The music generated in our experiments is used only for the analysis in this paper and is not released or distributed.

\section*{Generative AI Usage Statement}
A generative AI tool (Claude Opus 5.5) was used for language assistance, namely grammar checking and suggestions on wording.
It was also used to generate LaTeX code for formatting the results into tables and Python scripts for the figures.
All generated code and its outputs were reviewed by the authors.

\bibliographystyle{ACM-Reference-Format}
\bibliography{refs}

\appendix
\section{Experimental Details}
\label{app:details}

\subsection{Models and Fine-tuning Hyperparameter}
\label{app:finetuning}

\Cref{tab:app-finetuning} lists the fine-tuning recipe.
All four models, two sizes each of MusicGen~\cite{copet2023musicgen} and InspireMusic~\cite{zhang2025inspiremusic}, are fully fine-tuned on the same 30,000 ten-second segments with the same optimizer, schedule, effective batch size and number of steps.
MusicGen-L and InspireMusic-1.5B split each batch of 16 into smaller batches only to fit in memory, steps are accumulated to match as 16.
Each model is fine-tuned with three seeds, and the seed also fixes the data order.
For the deletion test, the removed training instances are taken out of this order without reshuffling, so the full fine-tuning, the random deletion and the targeted deletion with the same seed share the order of the remaining instances.

\begin{table}[!htbp]
  \caption{Models and fine-tuning recipe.}
  \footnotesize
  \centering
  \begin{tabularx}{\linewidth}{>{\raggedright\arraybackslash}p{2.6cm}>{\raggedright\arraybackslash}X>{\raggedright\arraybackslash}X>{\raggedright\arraybackslash}X>{\raggedright\arraybackslash}X}
    \toprule[1.2pt]
    Item & MusicGen-S & MusicGen-L & InspireMusic-Base & InspireMusic-1.5B \\
    \midrule
    Released model & facebook/musicgen-small & facebook/musicgen-large & iic/InspireMusic (Base) & iic/InspireMusic-1.5B \\
    \midrule
    Trained part & decoder (419.6M) & decoder (3,255.4M) & LLM (501.6M) & LLM (1,556.6M) \\
    \midrule
    Frozen & \multicolumn{2}{>{\hsize=\dimexpr2\hsize+2\tabcolsep\relax\raggedright\arraybackslash}X}{T5 text encoder~\cite{raffel2020exploring}, EnCodec~\cite{defossez2022high}} & \multicolumn{2}{>{\hsize=\dimexpr2\hsize+2\tabcolsep\relax\raggedright\arraybackslash}X}{WavTokenizer~\cite{ji2025wavtokenizer}, flow-matching model} \\
    \midrule
    Audio tokens per 10\,s & \multicolumn{2}{>{\hsize=\dimexpr2\hsize+2\tabcolsep\relax\raggedright\arraybackslash}X}{$4 \times 500$ (EnCodec, 32\,kHz, 50\,Hz)} & \multicolumn{2}{>{\hsize=\dimexpr2\hsize+2\tabcolsep\relax\raggedright\arraybackslash}X}{$1 \times 750$ (WavTokenizer, 24\,kHz, 75\,Hz)} \\
    \midrule
    Optimizer & \multicolumn{4}{l}{AdamW~\cite{loshchilov2019decoupled}, $\beta = (0.9, 0.95)$, weight decay 0} \\
    \midrule
    Learning rate & \multicolumn{4}{l}{$1 \times 10^{-5}$, 100 linear warmup steps, cosine decay to 0} \\
    \midrule
    Batch & 16 & $2 \times 8$ accumulation & 16 & $4 \times 4$ accumulation \\
    \midrule
    Steps & \multicolumn{4}{l}{7,500 (4 epochs)} \\
    \bottomrule[1.2pt]
  \end{tabularx}

  \label{tab:app-finetuning}
\end{table}

\subsection{TDA Methods Hyperparameter}
\label{app:tda}

\Cref{tab:app-tda} lists the hyperparameters of the TDA methods.

\begin{table}[!htbp]
  \caption{Hyperparameters of the TDA methods.}
  \footnotesize
  \centering
  \begin{tabularx}{\linewidth}{>{\raggedright\arraybackslash}p{2.6cm}>{\raggedright\arraybackslash}X}
    \toprule[1.2pt]
    Method & Hyperparameters \\
    \midrule
    GradDot & final checkpoint \\
    \midrule
    \GradCos~\cite{hanawa2021evaluation} & final checkpoint \\
    \midrule
    TracIn~\cite{pruthi2020estimating} & checkpoints at 1,875 / 3,750 / 5,625 / 7,500 steps, checkpoint weights $(1.73, 1.29, 0.59, 0.09) \times 10^{-2}$ \\
    \midrule
    \Trak~\cite{park2023trak} & projection dimension 8,192, $\lambda \in \{1, 0.1, 0.01\}$ \\
    \midrule
    \Trak ensemble & $\lambda = 1$, three fine-tuned models \\
    \midrule
    \LoGra~\cite{choe2024logra} & projection dimension 1,024 per layer, damping 0.1 \\
    \midrule
    \FactGraSS~\cite{hu2026grass} & projection dimension 1,024, intermediate dimension 4,096, damping 0.1 \\
    \bottomrule[1.2pt]
  \end{tabularx}

  \label{tab:app-tda}
\end{table}

\subsection{Clustering and Audio Features}
\label{app:features}

\Cref{tab:app-clustering} lists the features and the grouping of each criterion.
All $k$-means~\cite{lloyd1982least} clusterings use $k = 12$, the number of genres.
All audio features are computed with librosa~\cite{mcfee2015librosa} from the generated audio resampled to 48\,kHz, with an FFT size of 2,048 and a hop length of 512.

\begin{table}[H]
  \caption{Features and grouping of each criterion.}
  \footnotesize
  \centering
  \begin{tabularx}{\linewidth}{>{\raggedright\arraybackslash}p{2.6cm}>{\raggedright\arraybackslash}X>{\raggedright\arraybackslash}p{3.6cm}}
    \toprule[1.2pt]
    Criterion & Features of each generated output & Grouping \\
    \midrule
    TDA score clusters & $\mathbf{z}_t - \mathbf{z}^{\mathrm{glob}}$ over the 30,000 training instances, L2-normalized & $k$-means \\
    \midrule
    Genre & first tag of the prompt & 12 genres, 40 generated outputs each \\
    \midrule
    MFCC~\cite{davis1980mfcc} & mean and variance over time of 13 MFCCs (26 dimensions), standardized over the generated outputs & $k$-means \\
    \midrule
    Log-mel spectrogram~\cite{muller2015fundamentals} & mean over time of the log-mel spectrum (128 bands, dB), standardized over the generated outputs & $k$-means \\
    \midrule
    Chroma~\cite{muller2015fundamentals} & mean and variance over time of the CQT chroma (24 dimensions), standardized over the generated outputs & $k$-means \\
    \bottomrule[1.2pt]
  \end{tabularx}

  \label{tab:app-clustering}
\end{table}

\subsection{Computational Resources}
\label{app:compute}

\Cref{tab:app-compute} reports the GPU time of fine-tuning and of the deletion test.
The time is the elapsed time of each job on one GPU.
The retraining time is estimated as the median time per run multiplied by the number of runs.
Gradient computation, scoring, generation and measurement are not included.
In total, fine-tuning takes about 37 GPU hours and retraining for the deletion test about 830 GPU hours.

\begin{table}[H]
  \caption{GPU resources.}
  \small
  \centering
  \begin{tabularx}{\linewidth}{>{\raggedright\arraybackslash}p{1.9cm}>{\raggedright\arraybackslash}p{2.2cm}>{\raggedright\arraybackslash}X>{\raggedright\arraybackslash}X>{\raggedright\arraybackslash}X>{\raggedright\arraybackslash}X}
    \toprule[1.2pt]
    \multicolumn{2}{l}{Item} & MusicGen-S & MusicGen-L & InspireMusic-Base & InspireMusic-1.5B \\
    \midrule
    \multicolumn{2}{l}{GPU} & A40 & H100 & A40 & A40 \\
    \midrule
    \multirow{2}{=}{Fine-tuning} & time per run & 1.0\,h & 2.9\,h & 2.1\,h & 6.2\,h \\
    \cmidrule(l){2-6}
     & total (3 seeds) & 3\,h & 9\,h & 6\,h & 19\,h \\
    \midrule
    \multirow{3}{=}{Deletion test} & retraining runs & 153 & 55 & 153 & 51 \\
    \cmidrule(l){2-6}
     & time per run & 0.9\,h & 2.6\,h & 1.8\,h & 5.3\,h \\
    \cmidrule(l){2-6}
     & total & about 140\,h & about 145\,h & about 275\,h & about 270\,h \\
    \bottomrule[1.2pt]
  \end{tabularx}

  \label{tab:app-compute}
\end{table}

\section{Embedding Visualizations}
\label{app:embedding}

\begin{figure}[H]
  \centering
  \includegraphics[width=\linewidth]{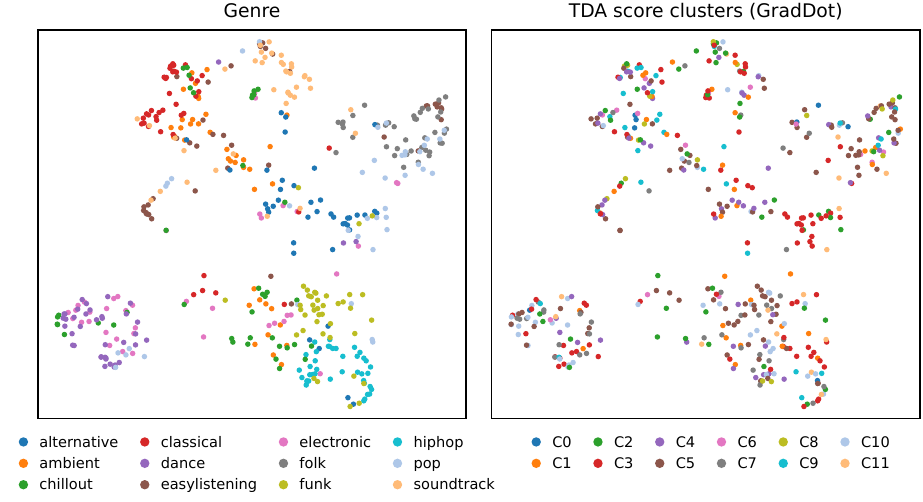}
  \caption{Generated outputs of MusicGen-S in the \MAEST embedding space, colored by genre (left) and by the TDA score clusters of GradDot (right).}
  \label{fig:embed-graddot}
\end{figure}

\begin{figure}[H]
  \centering
  \includegraphics[width=\linewidth]{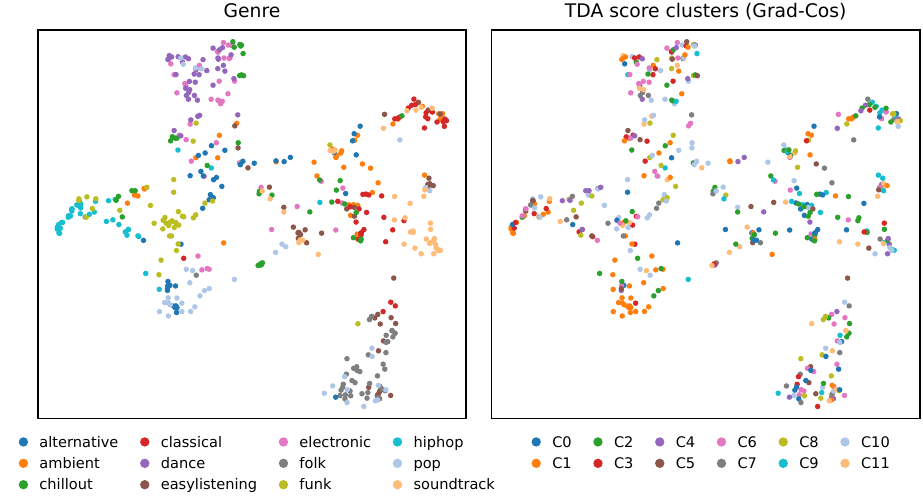}
  \caption{Generated outputs of MusicGen-L in the \MAEST embedding space, colored by genre (left) and by the TDA score clusters of \GradCos (right).}
  \label{fig:embed-gradcos}
\end{figure}

\begin{figure}[H]
  \centering
  \includegraphics[width=\linewidth]{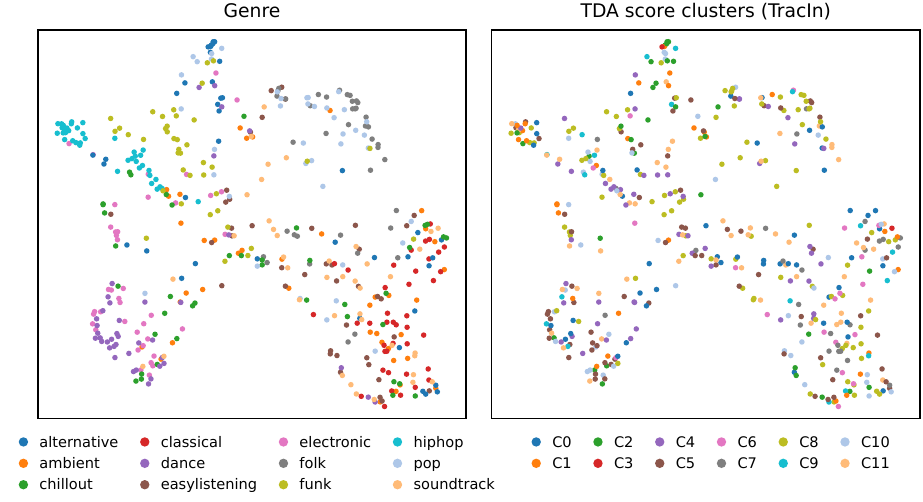}
  \caption{Generated outputs of InspireMusic-1.5B in the \MAEST embedding space, colored by genre (left) and by the TDA score clusters of TracIn (right).}
  \label{fig:embed-tracin}
\end{figure}

\begin{figure}[H]
  \centering
  \includegraphics[width=\linewidth]{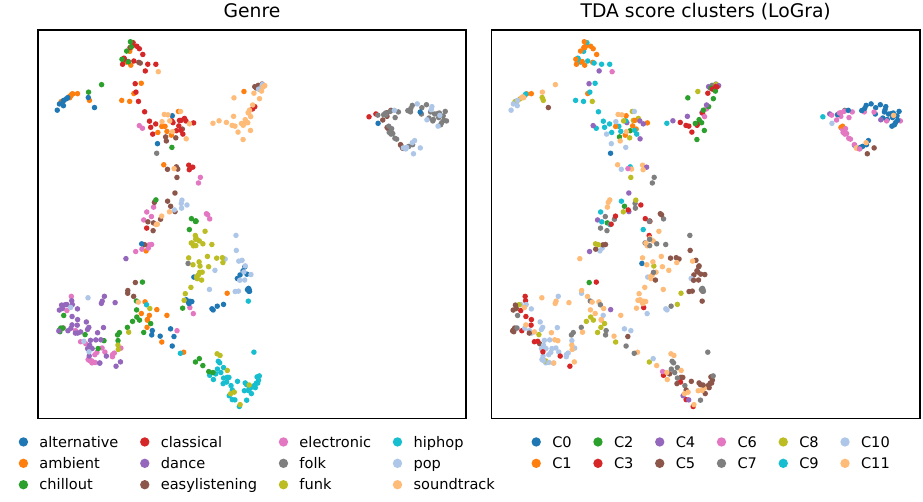}
  \caption{Generated outputs of MusicGen-L in the \CLAP embedding space, colored by genre (left) and by the TDA score clusters of \LoGra (right).}
  \label{fig:embed-logra}
\end{figure}

\begin{figure}[H]
  \centering
  \includegraphics[width=\linewidth]{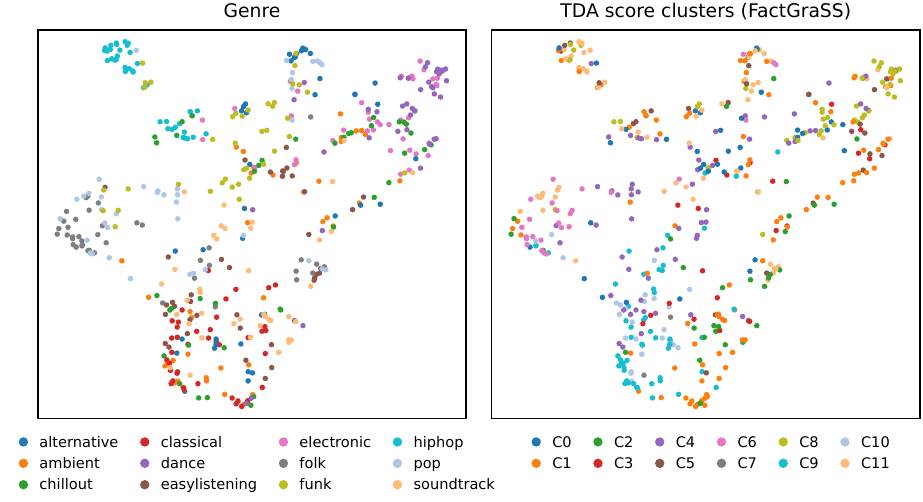}
  \caption{Generated outputs of InspireMusic-1.5B in the \CLAP embedding space, colored by genre (left) and by the TDA score clusters of \FactGraSS (right).}
  \label{fig:embed-factgrass}
\end{figure}

\Cref{fig:embed-graddot,fig:embed-gradcos,fig:embed-tracin,fig:embed-logra,fig:embed-factgrass} place the generated outputs of MusicGen-S, MusicGen-L and InspireMusic-1.5B in the \CLAP and \MAEST embedding spaces with UMAP~\cite{mcinnes2018umap}, and \cref{tab:app-embedding} lists the audio encoders.
In each figure, generated outputs of the same genre lie close to each other, whereas generated outputs of the same TDA score cluster are spread across the embedding space.

\begin{table}[!htbp]
  \caption{Audio encoders of the embedding visualizations.}
  \footnotesize
  \centering
  \begin{tabularx}{\linewidth}{>{\raggedright\arraybackslash}p{1.7cm}>{\raggedright\arraybackslash}p{4.6cm}>{\raggedright\arraybackslash}X}
    \toprule[1.2pt]
    Encoder & Checkpoint & Description \\
    \midrule
    \CLAP~\cite{wu2023large} & \nolinkurl{laion/larger_clap_music_and_speech} & audio encoder trained jointly with a text encoder by contrastive language-audio pretraining \\
    \midrule
    \MAEST~\cite{alonsojimenez2023efficient} & \nolinkurl{mtg-upf/discogs-maest-10s-pw-129e} & spectrogram-based audio transformer trained by multi-label music style classification \\
    \bottomrule[1.2pt]
  \end{tabularx}

  \label{tab:app-embedding}
\end{table}

\end{document}